\documentclass[twocolumn,floatfix,eqsecnum,floatfix,showpacs,elide]{revtex4-2}
\pdfoutput=1
\usepackage{graphicx}
\usepackage{float}
\usepackage[usenames,dvipsnames,svgnames,x11names]{pstricks}
\usepackage{amsmath}
\usepackage{amssymb}
\usepackage{fourier}

\newcommand{\REM}[1]{}

\usepackage{calc,tikz}
\definecolor{lime}{HTML}{A6CE39}
\DeclareRobustCommand{\orcidicon}{%
	\begin{tikzpicture}
	\draw[lime, fill=lime] (0,0) 
	circle [radius=0.16] 
	node[white] {{\fontfamily{qag}\selectfont \tiny ID}};
	\end{tikzpicture}
}

\usepackage{soul}
\sethlcolor{yellow}
\usepackage[ulem=normalem,authormarkup=none,truncate=hyphenate,todonotes={textsize=footnotesize,textwidth=1.4cm}]{changes}
\setaddedmarkup{\blue{#1}}
\setdeletedmarkup{\red\sout{#1}}
\definechangesauthor[color=yellow]{H}

\usepackage[%
unicode=true,%
colorlinks=true,%
linkcolor=blue,anchorcolor=blue,%
breaklinks=true,%
citecolor=blue,filecolor=cyan,%
menucolor=blue,%
urlcolor=blue]{hyperref}

\begin{document}

\title{
Control of a qubit state by a soliton propagating through a Heisenberg
spin chain}
\author{S. Varbev}
\author{I. Boradjiev}
\author{R. Kamburova}
\email{krad@issp.bas.bg}
\author{H.~Chamati\href{https://orcid.org/0000-0002-0831-6945}{\orcidicon}\hspace*{-0.2cm},}
\email{chamati@issp.bas.bg}
\homepage[\orcidicon]{https://orcid.org/0000-0002-0831-6945}
\affiliation{Institute of Solid State Physics, Bulgarian Academy of Sciences, Tzarigradsko chauss\'{e}e 72, 1784 Sofia, Bulgaria}

\date{\today}

\begin{abstract}
We demonstrate that nonlinear magnetic solitary excitations -- 
solitons -- traveling
through a Heisenberg spin chain may be used as a robust tool capable
of coherent control of the qubit's state.
The physical problem is described by a Hamiltonian
involving the interaction between the soliton and the qubit. We show
that under certain conditions the generic Hamiltonian may be mapped on that
of a qubit two-level system with matrix elements depending on the
soliton parameters. We considered the action of a bright and a dark
solitons depending on the driving nonlinear wave function.
We considered a local interaction restricted the closest to the qubit
spin
in the chain. We computed the expressions of the physical
quantities of interest for all cases
and analyzed their behavior in some special limits.
\end{abstract}


\maketitle

\section{Introduction}
The control of spin dynamics is a prerequisite for effective
quantum computing and quantum information processing
\cite{nielsen_quantum_2010,marinescu_classical_2012,band_quantum_2013,grumbling_quantum_2019,gaita-arino_molecular_2019,atzori_second_2019}. The research in this field is focused
on the physical implementation
of the quantum computer, that relies on the realization of a two-level
quantum system -- the qubit. Over the last few decades, a number of two-level
systems have enjoyed great interest as potential candidates for qubits. These
include among other, the
electrons confined in quantum dots due to their very long dephasing
times and long phase coherence lengths \cite{Engel_2004},
nanometer-scale magnetic particles or clusters with large total spin
and high anisotropy \cite{Tejada_2001}, molecular magnets consisting
of clusters with coupled transition metal ions
\cite{PhysRevLett.98.057201} and a transmon-type superconducting
qubit \cite{tabuchi_coherent_2015}. 

To manipulate the state of qubits, there are different approaches involving
external stimuli,
such as an electric field, a laser beam, a magnetic field or microwave
pulses. The properties of systems composed
of a qubit under the action of one of these stimuli have been the subject of
extensive research (see Refs.
\cite{marinescu_classical_2012,tabuchi_coherent_2015,chechik_molecules_2016,atzori_second_2019,wasielewski_exploiting_2020,froning_ultrafast_2021,froning_ultrafast_2021,atzori_radiofrequency_2021}
and references therein). On the other hand, a scarcely explored path to
alter the state of a qubit takes account of
an effective well localized in space and time solitary wave
propagating through a quantum spin chain -- the soliton. Solitons
are collective excitations that emerge in a variety of nonlinear systems spanning all
branches of physics, such as optical fibers, cold atoms, fluids, magnets and
so on. The study of nonlinear spin dynamics in magnetic materials
related to different phenomena in condensed matter physics has
been the subject of considerable interest for decades and a series of
theoretical investigations
\cite{akhiezer_theory_1967,tjon_solitons_1977,gochev_spin_1977,kosevich_nonlinear_1977,pushkarov_solitary_1977,perelomov_generalized_1977,huang_soliton_1990,rakhmanova_intrinsic_1998,ivanov_effective_2003,stancil_spin_2009,primatarowa_dark_2012,latha_integrable_2014,bulut_dark_2018,jeyaseeli_intrinsic_2021}
has concentrated on unveiling the underlying
mechanism of soliton formation in the framework of different magnetic models
(for reviews the interested reader may consult Refs.
\cite{kosevich_magnetic_1990,mikeska_solitary_1991,lai_nonlinear_1999,kalinikos_nonlinear_2013}
and references therein). The ever growing
interest in solitary spin waves that can travel under certain
physical conditions with constant shape and velocity is also due to
their applications in various technological fields, such as
spintronics \cite{hirohata_review_2020}.
In this respect, there are only a few papers dealing with the
control of the qubit by a
soliton excitations propagating along a Heisenberg spin chain
\cite{Cuccoli_2014_1,Cuccoli_2014_2,Cuccoli_2015,Varbev_2019,Varbev_2021}.
In this system, a solitary wave may be coherently generated by exciting the
spin at one end of the chain via a time-dependent magnetic field.
On the experimental side the potential coherent manipulation of
a transmon--type superconducting qubit with the aid of a ferromagnetic
single--magnon excitation in a millimeter-sized ferromagnetic sphere
was considered in Ref. \cite{tabuchi_coherent_2015}.

To explore the physics of a system consisting of an interacting
soliton with a qubit, the traveling soliton is generated
a distance apart from the qubit in order to reduce noise effects that might
break quantum coherence. Moreover, the soliton parameters
and the strength of the qubit-chain coupling may be tuned to control the qubit states.
The qubit dynamics governed by a soliton propagating in a magnetic
chain was numerically studied with the aid of the Bloch equations in Refs.
\cite{Varbev_2019,Varbev_2021}. There, the chain was modeled by an anisotropic
Heisenberg model with spin-spin interaction restricted either to
nearest-neighbors or extended to include next to nearest-neighbors,
assuming that the qubit does not affect the stability of the soliton, while
the soliton may be used to control the behavior of the qubit. In the
present paper, we tackle this problem in a more rigorous fashion and
show that this assumption is indeed valid when the magnitude of the
spin in the chain is sufficiently large. Hence, we demonstrate
that in the large-spin limit the quantum problem reduces to the
corresponding nonlinear Schr\"{o}dinger equation for a two-level quantum system
\cite{band_quantum_2013} for the qubit in the
quasi-classical approximation, i.e. when
the magnetic soliton stands for a large number of magnons \cite{kosevich_magnetic_1990}. We would like
to anticipate that unlike familiar two-level quantum systems, the Hamiltonian
derived here has a peculiar property that is
all its matrix elements depend in a complex way on the
soliton's characteristics. On the other hand,
the soliton is obtained as a solution of quasi-classical
equations of motion for the spin chain described by the Heisenberg
model with nearest-neighbor spin-spin interaction.
Then, we proceed with the derivation of the approximated analytic
solutions for the qubit, and find the probabilities for qubit
switching.
Our analysis of the physical problem shows that by a suitable choice of the parameters of the
soliton it is possible to control the dynamics of the qubit.

This paper is organized as follows:
The quantum model of interest is defined in Section \ref{s-q}.
In Section \ref{semi}, we demonstrate how the original problem
reduces to the study of a quasi-classical two-level system.
The effect of the quasi-classical solitons and possible ways to control the
qubit state are analyzed in Section \ref{Sol-Drive}. To achieve our goal, 
we consider
a local coupling of the qubit and its closest spin on the chain to
study the effects of two types of solitons
-- bright and dark -- on the behavior of the qubit. In the last Section \ref{Sec:Discussion_and_Conclusions}, we
summarize our results.

\section{the Spin chain -- Qubit system}\label{s-q}
The system of interest is composed of a qubit
anisotropically interacting with an anisotropic Heisenberg spin chain.
A general expression of the Hamiltonian reads
\begin{align}
H = & -J\sum_{n}\mathbf{S}_{n}\cdot\mathbf{S}_{n+1} 
-A \sum_{n}(S^{z}_{n})^2 -\mu H_{0}\sum_n S^z_n \nonumber \\
&
- d_{xy}
\left[{S^x_{0}}{\sigma}^{x}+{S^y_{0}}{\sigma}^{y}\right]
- d_{z}{S^z_{0}}{\sigma}^{z} - \nu H_{0}\sigma^{z}
\label{ModelHamiltonian}
\end{align}
with $\mathbf{S}_{n}\equiv(S_{n}^x,S_{n}^y,
S_{n}^z)$ -- the spin operator located at the site $n$,
$\boldsymbol\sigma\equiv\left(\sigma^x,\sigma^y,\sigma^z\right)$
is the half-spin operator associated to the qubit, to be dubbed hereafter
``qubit'' for short. The constants
$J>0$ and $A>0$ stand for the spin coupling and
the strength of the crystal field anisotropy, respectively. $H_0$ is an external magnetic field
oriented along the $z$-direction.
$\mu$ and $\nu$ are the magnetic moments per spin
in the chain's sites
and the qubit, respectively. The interaction between the qubit
and the spin chain is parametrized by the coupling's components
$d_{xy}$ within the $xy$ plane and $d_z$ along the $z-$axis.
Here and below, we set $\hbar=1$, i.e. we work in units of $\hbar$ and
assume that the qubit is coupled to the spin sitting at
site $n=0$ of a long (quasi-infinite) spin chain with periodic
boundary conditions, thus minimizing
the effect of the boundaries on the properties of the chain.

It is convenient to express Hamiltonian
(\ref{ModelHamiltonian}) in terms of raising and lowering spin
operators designated by ``$+$'' and ``$-$'', respectively, i.e.
$$
S_n^{\pm} = S^{x}_n \pm i S_n^{y} \quad \mathrm{and} \quad
\sigma^{\pm} = \sigma^{x} \pm i \sigma^{y},
$$
that obey the conventional spin commutation relations.
Thus, it can be easily seen that the ensuing Hamiltonian possesses an
invariant that is the $z$-projection
of the total spin
$\boldsymbol{\Sigma}=\mathbf{S}+\boldsymbol{\sigma}$ of the physical
system -- spin chain and qubit, i.e.
$$
\Sigma^z = S^{z} + {\sigma}^{z}.
$$
It is associated to the global $U(1)$ gauge symmetry $S^{\pm}_n \to
S^{\pm}_n e^{\pm i\phi}$ and $\sigma^{\pm} \to
\sigma^{\pm}e^{\pm i\phi}$, where $\phi$ is an arbitrary real
constant.
Thus, by adding and subtracting the term $\mu H_0 \sigma^z$ to Hamiltonian
\eqref{ModelHamiltonian}, we end up with the invariant structure $-\mu H_0\Sigma^z$,
which just shifts the energy scale, and after a subsequent phase
transformation, it adds up to the phase of the state vector. Finally,
the resulting Hamiltonian
may be split into two components
\begin{subequations}
\begin{align}
\label{Hamiltonian_Tr}
H+\mu H_0\Sigma^z = H^{\text{c}} +
H^{\text{q}} ,
\end{align}
where
\begin{equation}
\label{Hamiltonian_c}
H^{\text{c}} = -J \sum_{n} \mathbf{S}_{n}\cdot\mathbf{S}_{n+1} - A\sum_{n}(S^{z}_{n})^2  
\end{equation}
describes the magnetic properties of the Heisenberg spin
chain in the absence of the external field, and
\begin{equation}
\label{Hamiltonian_q}
H^{\text{q}} = \left[(\mu-\nu)H_0 - d_{z}{S^z_{0}}\right] \sigma^{z}
- \frac{d_{xy}}{2} \left({S^+_{0}}{\sigma}^{-}+{S^-_{0}}{\sigma}^{+}\right)
\end{equation}
\end{subequations}
is associated to the qubit and its interactions with the external
field $H_0$, as well as the chain.

Our main task here is to coherently modulate the
qubit state through manipulation of the state of the spin chain.
Thus, we find it more appropriate to work in the framework of the
Heisenberg picture
with respect to Hamiltonian \eqref{Hamiltonian_c}
using the transformation
\begin{subequations}
\begin{align}\label{Hamiltonian_t}
\vert\psi(t)\rangle &\to e^{iH^\text{c} t} \vert\psi(t)\rangle, \\
H^\mathrm{q} &\to e^{iH^\text{c} t} H^\text{q} \,e^{-iH^\text{c} t} , 
\end{align}
\end{subequations}
where $\vert\psi(t)\rangle$ stands for the time dependent state
vector. Then,
the transformed Hamiltonian takes the expression
(\ref{Hamiltonian_q}) by replacing
the spin $\mathbf{S}_{0}$ with its time dependent counterpart.

\section{quasi-classical Approximation}\label{semi}
The equations of motion of a spin sitting on the Heisenberg
chain are given by
\begin{equation}
\label{EOM}
i \frac{\partial}{\partial t}\mathbf{S}_n=[\mathbf{S}_n,H^c].
\end{equation}
For the ladder operators, we have
\begin{align}
\label{EOM1}
\pm i \frac{\partial}{\partial t}S^\pm_{n}= &
	-J\left[S^z_n\left(S^\pm_{n-1}+S^{\pm}_{n+1}\right) -
S^\pm_n\left(S^z_{n-1}+S^z_{n+1}\right)\right] \nonumber\\
&+ A\left(S^{z}_{n}S^{\pm}_{n} +
	S^{\pm}_{n}S^{z}_{n}\right).
\end{align}

For the purpose of this paper, we assume that the quantum
number $S$ is large enough to consider the quasi--classical limit
\cite{dauxois_physics_2010}.
Therefore, we will work in the large-$S$ approximation where the
physical properties of the spin chain obey the laws of classical
mechanics. In this case the back-action of the qubit on the chain may
be neglected. Such regime can always be achieved
in the large-$S$ limit due to the fact that the spin--chain component
$H^c$ scales as $S^2$ while the qubit component $H^q$ like $S$.
Moreover, we assume that the deviation of the spin
from its maximal projection, $S_n^z = S$, along the $z$-axis is small enough.

Within this quasi-classical approximation, the components of the spin
operators transform into classical functions:
$S_{n}^+=S\alpha_n$, $S_{n}^-=S\alpha_n^*$,
$S_{n}^z=S\sqrt{1-|\alpha_n|^2}$.
Substituting in \eqref{EOM1},
we get
\begin{align}
\label{EOMa}
i\frac{\partial \alpha_n}{\partial t} =\ & 2AS
\alpha_{n}\sqrt{1-|\alpha_{n}|^{2}}
	\nonumber\\
& - JS
	\left[(\alpha_{n+1}+\alpha_{n-1})\sqrt{1-|\alpha_{n}|^{2}}\right.
	\notag \\
	& \ \left.\qquad - \alpha_{n}
\left(\sqrt{1-|\alpha_{n+1}|^{2}}+\sqrt{1-|\alpha_{n-1}|^{2}}\right)\right].
\end{align}
A similar equation is obtained for the corresponding complex conjugate
$\alpha^*$. 
To solve \eqref{EOMa}, we look for solutions in the form of
amplitude--modulated waves, i.e.
\begin{equation}
\alpha_{n}(t) = \varphi_n(t)e^{\textstyle i(k n-\omega t)},
\end{equation}
where $\varphi_n(t)$ denotes the envelope of $\alpha_{n}(t)$, $k$ --
the wave number, $\omega$ -- the angular frequency of the carrier
wave. Notice that the phase of the functions $\alpha_n(t)$ is linear
in both the position of the
spin and time by virtue of the approximation of small spin deviations from
their maximal projection along the $z$-axis.

For further analysis, we
employ the semidiscrete approach
\cite{flytzanis_kink_1985,pnevmatikos_soliton_1986,remoissenet_low-amplitude_1986,primatarowa_exciton_1995},
\textit{i.e.} we treat the
phase factor exactly and use a continuum approximation for the
envelope $\varphi_n(t)$. This removes the restriction for long carrier
wavelengths and permits the study of envelope solitons with
arbitrary wave numbers inside the Brillouin zone.

In the
continuum limit, corresponding to a vanishing distance, say $\Delta x$, between
nearest-neighbor spins on the chain with a large number of spins $N$, i.e. in the limit
$\Delta x \to 0$ and $N\to \infty$, provided $N\Delta x$ remains finite,
the discrete function $\varphi_n(t)$ may be approximated by
the smoothly varying time dependent real function $\varphi(x,t)$ describing the deviation
of the spin, located at a position $x$, with respect to
the $z-$axis.
Then, for wide excitations with width, $L \gg \Delta x$, and an envelope satisfying
\begin{equation}
[\varphi(x,t)]^2 \ll 1,
\label{SmallDeviation_Approximation}
\end{equation}
Eq. \eqref{EOMa} transforms into the nonlinear Schr\"{o}dinger equation (NLSE)
\begin{equation}
i\left(\frac{\partial\varphi}{\partial t} + v_g \frac{\partial\varphi}{\partial x}\right) = (\omega_0 -\omega) \varphi  - b_kS\frac{\partial^2\varphi}{\partial x^2} +
g_kS|\varphi|^{2}\varphi ,
\label{NLSE}
\end{equation}
with $\omega_0 =$ $-2g_kS$ -- the characteristic frequency of the magnon,
$v_g =2SJ \sin k$ -- the group velocity of the carrier wave,
$b_k=J \cos k$ -- the dispersion coefficient, and
$g_k=J(\cos k - 1)- A$ -- a nonlinear coefficient that
incorporates the spin exchange and anisotropy interactions that are both nonlinear.
It is worth mentioning that the sign of the quantity $b_kg_k$, that is
a function of the anisotropy constant $A$ and the wave number $k$, determines
the nature of the soliton solutions of NLSE \eqref{NLSE}. Thus, we
have bright solitons for negative values ($b_kg_k<0)$, while
for positive values ($b_kg_k>0$) dark solitons may be achieved.
Bright and dark solitons, as exact solutions of the NLSE
\eqref{NLSE}, have been studied intensively over the last few decades
\cite{novikov_theory_1984,kivshar_dark_1998,ablowitz_nonlinear_2011,agrawal_nonlinear_2019}.
Notice that, the ensuing soliton propagates, with a velocity $v_g$,
over the distance $L$, during the period $T=\frac{L}{v_g}$.
Below, we will investigate numerically
the stability of the soliton structures,
with the aid of the predictor-corrector algorithm. Moreover, the
accuracy of the computations is controlled through the
conservation of the envelope amplitude squared.
More details on numerical techniques for the
investigating the properties of bright and dark solitons can be found in Ref.
\cite{bao_numerical_2013}.

Within our quasi-classical approach in the Heisenberg picture complemented with the phase transformation
\begin{align}
e^{ i (kx-\omega t)\,\sigma^z},
\label{phase_transformation_gamma}
\end{align} 
the time dependent Hamiltonian \eqref{Hamiltonian_q} reduces to the
effective two-state Hamiltonian, given by
\begin{subequations}
\begin{equation}
H^q(t) = \frac12\Omega(t)
\left(\sigma^{-} + {\sigma}^{+} \right)
+ \Delta(t) \, \sigma^{z},
\label{Hamiltonian_q_Approximated} 
\end{equation}
where
\begin{equation}
\Omega(t) = - d_{xy} S \varphi(x,t) \label{Omega_def}
\end{equation}
is the time-dependent coupling and
\begin{equation}
\Delta(t) = (\mu-\nu) H_0 -d_{z} S\sqrt{1 -
\left[\varphi(x,t)\right]^2} +\omega,
\label{delta_def} 
\end{equation}
is the time-dependent detuning.
\end{subequations}

Within the above mentioned quasi-classical approach, the action of the soliton on
the qubit is fully embodied in the parameters
$\Omega(t)$ \eqref{Omega_def} and $\Delta(t)$ {\eqref{delta_def}}. From the physical
point of view, Hamiltonian {\eqref{Hamiltonian_q_Approximated}}
represents a two-level system perturbed under the action of the
soliton. Here the coupling $\Omega(t)$ is related to the transition between both
levels and the detuning $\Delta(t)$ stands for the difference in
energy between these levels. It is worth noticing that
the coupling $\Omega(t)$ is proportional to the in-plane anisotropic
parameter $d_{xy}$ times the envelope $\varphi(x,t)$, while the detuning
$\Delta(t)$ is a linear function of the anisotropic coupling $d_z$,
the external magnetic field, as well as the frequency of the carrier
wave.

To explore the dynamics of the effective Hamiltonian
\eqref{Hamiltonian_q_Approximated}, we work in the standard spin basis
of the qubit, starting from the Schr\"odinger equation and after some
lengthy but straightforward algebra, we end
up with the Schr\"odinger equation for
the probability amplitudes $C_{\pm}(s)$, associated to the states $\left\vert \pm
\tfrac12\right\rangle$, given by
\begin{equation}
i \frac{\text{d}}{\text{d}s} 
\begin{bmatrix}
C_{-}(s)\\
C_{+}(s)
\end{bmatrix}
=
\begin{bmatrix}
-\Theta(s) & 1 \\
1 & \Theta(s)
\end{bmatrix} 
\begin{bmatrix} 
C_{-}(s) \\
C_{+}(s)
\end{bmatrix} .
\label{Schrodinger_s}
\end{equation}
Here, we follow a standard procedure commonly used in solving
the dynamical problem of two-level systems
\cite{delos_solution_1972,crothers_semiclassical_2008}. It consists of
finding a solution of
the time-dependent equations describing the evolution of the effective Hamiltonian
\eqref{Hamiltonian_q_Approximated} with the aid of the
reduced-time variable
\begin{equation}\label{st}
s(t) = \frac{1}{2} \int_{0}^{t} \Omega(u)\, \text{d}u ,
\end{equation}
and the St\"uckelberg variable
\begin{equation}\label{thetas}
\Theta(s) = \frac{\Delta[t(s)]}{\Omega[t(s)]}
\end{equation}
with initial conditions
$$
C_-[s(-t_i)] = 0, \quad C_+[s(-t_i)] = 1
$$
i.e. at $t= -t_i$, the qubit points in the $z$-direction.

We find this description of the considered two-level system problem
more convenient, since a single $s$-dependent quantity in
the Hamiltonian matrix $H(s)$ will make the subsequent analysis more
transparent, although it does not fully decouple the
Schr\"odinger equation \eqref{Schrodinger_s}.

\section{Soliton-Driven Qubit Dynamics}\label{Sol-Drive}
We examine the effect of the two different sorts of solitons --
bright and dark -- propagating through an anisotropic Heisenberg spin chain on
the qubit a distance apart from the spins.
So far, we established that
the corresponding general problem reduces to the two-level system problem
for the qubit (\ref{Schrodinger_s}), where the soliton
state characteristics are encoded in the variables \eqref{st} and \eqref{thetas} through
the coupling \eqref{Omega_def} and the detuning \eqref{delta_def}.

\subsection{Bright--soliton drive}
When $b_kg_k < 0$, for an infinite chain subjected to the
boundary conditions $|\varphi_b(x,t)|^2\rightarrow
0$ as $x\rightarrow\pm \infty$, the solution of NLSE \eqref{NLSE}
is a bright soliton given by
\begin{subequations}
\begin{equation}
\label{bs}
\varphi_b(x,t) = \varphi_{b0} \, {\rm sech}\, \frac{x-vt}{L}
\end{equation}
with amplitude
\begin{equation}
\varphi_{b0} = \frac{1}{L}\sqrt{\frac{2J\cos k}{J(1-\cos k)+A}},
\end{equation}
and frequency
\begin{equation}
\omega_b = 
\omega_0- 2SJ\frac{\cos k}{L^2}.
\end{equation}
\end{subequations}
That is, the soliton frequency is shifted relative to the linear spin wave
frequency $\omega_0$. Further, we may choose the wave number $k$ and
soliton width $L$ ($L\gg\Delta x$) as running soliton parameters of
both bright and dark solitons. The domain of existence of the bright soliton
solution \eqref{bs} in the case of an easy axis anisotropy
($A>0$), is $0\leq k<\tfrac{\pi}{2}$. Notice that the soliton is at rest
when $k=v=0$. Bright soliton solutions expressed by \eqref{bs} for a
one-dimensional system of classical spins with nearest neighbour
Heisenberg interaction were obtained and analyzed in details in Refs.
\cite{lakshmanan_continuum_1977,wallis_intrinsic_1995}.

\begin{figure}[h!]
\centering
\includegraphics[width=\columnwidth]{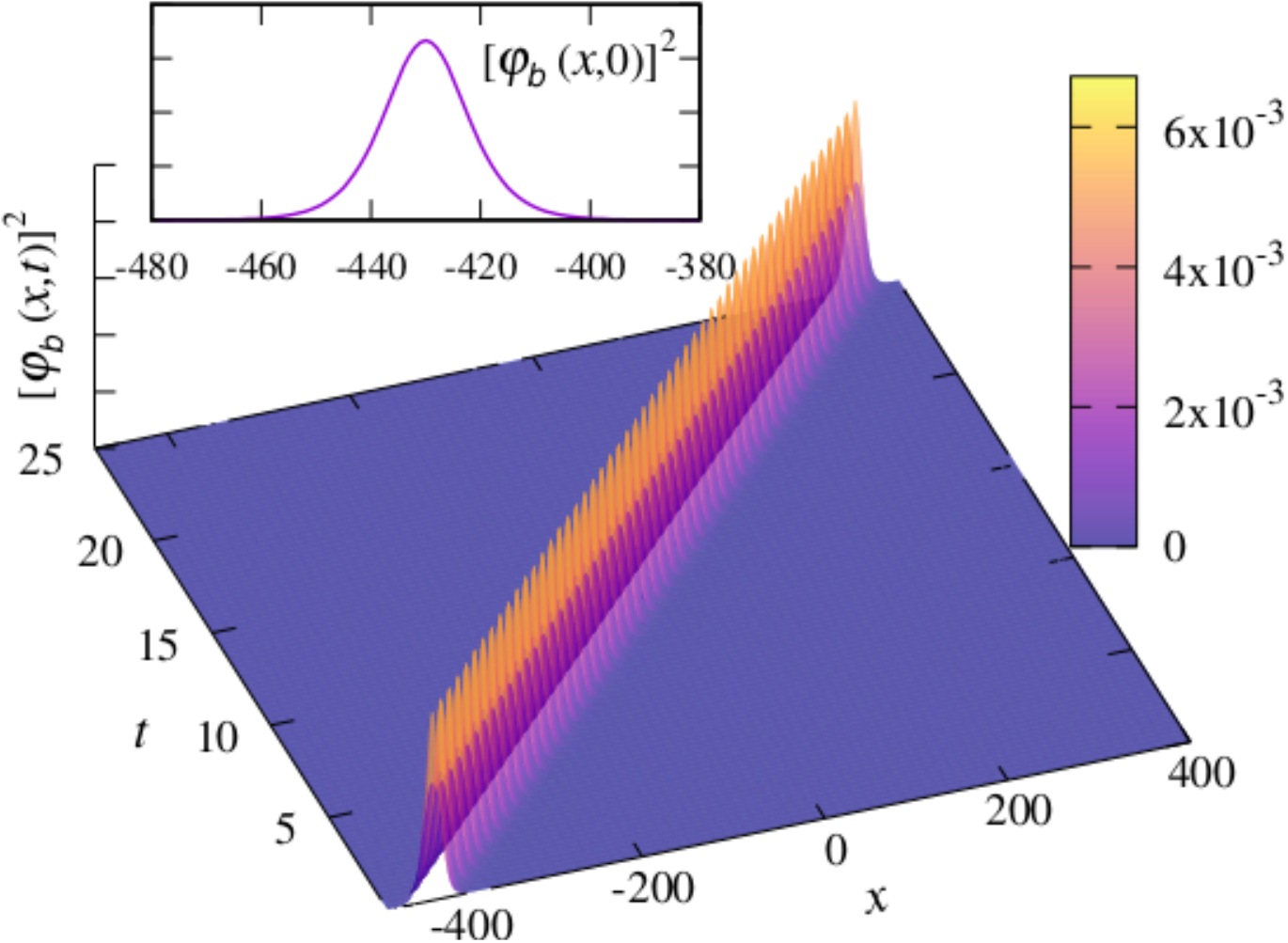}
\caption{Time evolution of the square of the bright soliton solution,
$[\varphi_b(x,t)]^2$,
given by \eqref{bs} using the parameters $k = 0.015708$ and
$L=10$ for $S=J=1$ and $A=3$. The numerical analysis demonstrates that the soliton solution remains
stable over time while propagating throughout the ferromagnetic
Heisenberg spin chain. In
the inset, we show the shape of the solitary wave as a function of the
position on the spin chain at the initial time $t=0$.}
\label{brightsol}
\end{figure}

The propagation of a bright soliton generated at initial time $t = 0$
away from the qubit position has been investigated via numerically
solving the system of discrete equations of motion \eqref{EOMa}. We used a
chain composed of 1000 spin sites, $J= S=1$, $A=3$ and
periodic boundary conditions (see Fig. \ref{brightsol}). It is easily seen that the
bright solitary wave remains stable during its evolution in time
throughout the spin chain.

For a bright soliton propagating in the chain and a qubit coupled
to a single chain spin, the variables \eqref{st} and \eqref{thetas}
are explicitly given by
\begin{subequations}
\begin{equation}
	s_b(t) = \Omega_{b0}T\left( \arctan e^{\frac tT} -\frac{\pi}{4}\right) , 
\end{equation}
and
\begin{equation}
	\Theta_b(s) = - \frac{\Delta_{b0}}{d_{xy} S} \,\, \tilde{\varphi}_b(s)^{-1}
- \frac{d_z}{2d_{xy}} \,\, \tilde{\varphi}_b(s) , 
\label{Theta_BS-SSD}
\end{equation}
where the amplitude of the coupling \eqref{Omega_def} is now expressed by
\begin{equation}
\Omega_{b0} = - d_{xy}S\varphi_{b0}, \label{Omega_BSSQ}
\end{equation}
the time-independent term of the detuning \eqref{delta_def} reads
\begin{equation}
\Delta_{b0} = 
(\mu-\nu)H_0-d_zS+\omega_b
\label{delta_BSSQ}
\end{equation}
and
\begin{equation}
\tilde{\varphi}_b(s) = \varphi_b[t(s)] = \varphi_{b0} \cos\left(\frac{\pi}{2} \frac{s}{s_{\infty}}\right).
\end{equation}
\end{subequations}
In this case, we placed the qubit at $x=0$, and expanded the square root in the detuning
(\ref{delta_def}) in Taylor series up to the first order with respect
to the function $\tilde{\varphi}_b(s)^2\propto\varphi_{b0}^2 \ll1$.
This implies that when 
\begin{equation}
\vert\Delta_{b0}\vert \gg \frac{1}{2} \, \vert d_z\vert S \varphi_{b0}^2,
\label{delta_large_BS-SSD}
\end{equation}
we can safely neglect the second term in (\ref{Theta_BS-SSD}) to end
up with
\begin{equation}
\Theta_b(s) =
\frac{\Delta_{b0}}{\Omega_{b0}}\,{\rm sec} \left(\frac{\pi}{2} \frac{s}{s_{\infty}} \right),
\end{equation}
that is identical to the St\"uckelberg variable of the Rosen-Zener model
\cite{Rosen1932} with coupling 
$\Omega_{b0}\mathrm{sech}\left(\tfrac{vt}L\right)$
and time-independent
detuning (\ref{delta_BSSQ}). Thus, within this approximation there is
a direct mapping of our model onto Rosen--Zener's one.
Then, the solution for the final time spin-flip probability is given by
\begin{equation}
P_{-} = P_- (+\infty) \approx \frac{\sin^2 \left( \frac{\pi}{2}
\Omega_{b0} T \right) } {\cosh^2 \left( \frac{\pi}{2}\Delta_{b0} T \right) } .
\end{equation}
Whence the transition probability exhibits oscillations with amplitude
$\cosh^{-2} \left( \frac12\pi \Delta_{b0} T \right)$ as a function of
the zero--mode interaction
$
\int_{-\infty}^\infty\Omega_b(t)e^{-i0 t}dt=\pi\Omega_{b0}T.
$
With the proper choice of control parameters, one
may create superposition states satisfying the inequality
$$
\vert C_-(+\infty) \vert
\cosh \left( \frac12\pi \Delta_{b0} T \right)\leq 1,
$$
given that condition
(\ref{delta_large_BS-SSD}) is fulfilled. However, a qubit-flip for
$\Delta_{b0} \neq 0$ is not possible since the amplitude
$C_-(+\infty)$ decays exponentially as a function of $\Delta_{b0} T$.

Remark that a control over the magnitude of the transition
probability can be achieved on-resonance, i.e. when
\begin{equation}
	\Delta_b(t) = 0  ,
\label{delta_resonance_BS_SSQ}
\end{equation}
then, we have
\begin{equation}
	P_{-} = \sin^2 \left( \frac{\pi}{2} \, \Omega_{b0} T \right)  . 
\label{probability_On-resonance}
\end{equation}
Since the detuning is a time-dependent quantity, while we assume that
the tuning parameters are time-independent, the resonance condition
(\ref{delta_resonance_BS_SSQ}) cannot be satisfied. Therefore, we
shall rather seek to achieve an effective on-resonance
regime. This requires to take into account the fact that the term
$\tilde{\varphi}_b(s)^2$ cannot be neglected and
(\ref{delta_large_BS-SSD}) does not need to be fulfilled.
To proceed further, we choose the coupling $d_z$ to be a tunable parameter
and split it into two components
\begin{subequations}\label{coupling_d_z}
\begin{equation}
d_z = d_{z0} + d_{z1}, 
\end{equation}
with the time-independent quantity
\begin{equation}
d_{z0} = \frac{1}{S}\left[\omega_b + (\mu-\nu) H_0 \right] , \label{coupling_d_z0} 
\end{equation}
obtained by setting $\Delta_{b0}\equiv\lim_{t\to-\infty}\Delta_b(t)=0$. The
detuning time dependence, on the other hand,
is effectively taken care of via the expression
\begin{equation}
d_{z1} = \frac{d_{z0}}{2} \, \left(\frac{\varphi_{b0}}{\eta}\right)^2,
\label{coupling_d_z1}
\end{equation}
\end{subequations}
where $\eta$ is a suitably chosen dimesionless
running parameter of the order of unity.
Expression (\ref{coupling_d_z1}) approximately
fulfils (\ref{delta_resonance_BS_SSQ}), which holds only during the time
period
(of the order of $T$) when the soliton interacts with the qubit, i.e.
when a transition
takes place. At early and late times, in the absence of
interaction, the system is slightly detuned, altering nothing but the
phases of the amplitudes. The ``relaxed'' requirement (\ref{coupling_d_z})
provides a more accurate approximation to the on-resonance regime
(\ref{delta_resonance_BS_SSQ}) than the strict setting $\Delta_b = 0$,
moreover it allows the use of the transition probability formula
(\ref{probability_On-resonance}).

In general, the effective resonant coherent control can be accomplished
by demanding that
\begin{subequations}\label{probability_xi_BS_SSQ}
\begin{equation}
P_{-} = \xi, \quad 0 \leq \xi \leq 1 ,
\end{equation}
yielding
\begin{equation}
	\pi\Omega_{b0} T = \pm \, 2 \arcsin \sqrt{\xi} - 2p\pi, \quad p \in
\mathbb{Z},
\label{omega_xi_BS_SSQ}
\end{equation}
\end{subequations}
for the zero-mode interaction.
In particular, in order to achieve properties,
such as complete population transfer (qubit switching)
$P_{-} = 1$, complete population return (qubit switching with
consecutive qubit return) $P_{-} = 0$, and equal
superposition $P_{-} = \frac 12$, we require that the zero-mode
interaction fulfills
$\pi\Omega_{b0} T = (2p+1)\pi, \, 2p\pi$ $(p\neq 0)$, and
$\left(p+\frac12\right)\pi$,
where $p \in \mathbb{Z}$, respectively.

To ensure the fulfillment of (\ref{probability_xi_BS_SSQ}) we can
choose the coupling $d_{xy}$ to be a tunable parameter, that satisfies
\eqref{Omega_BSSQ}
with $\Omega_{b0}$ from (\ref{omega_xi_BS_SSQ}). 
To summarize, the manipulation of the values of the spin chain -- qubit
coupling components $d_z$ (\ref{coupling_d_z}) and $d_{xy}$ \eqref{Omega_BSSQ}
at given soliton parameters allows to establish an effective
on-resonance coherent control over the state of the qubit.

\begin{figure}[h!]
\centering              
\includegraphics[width=\columnwidth]{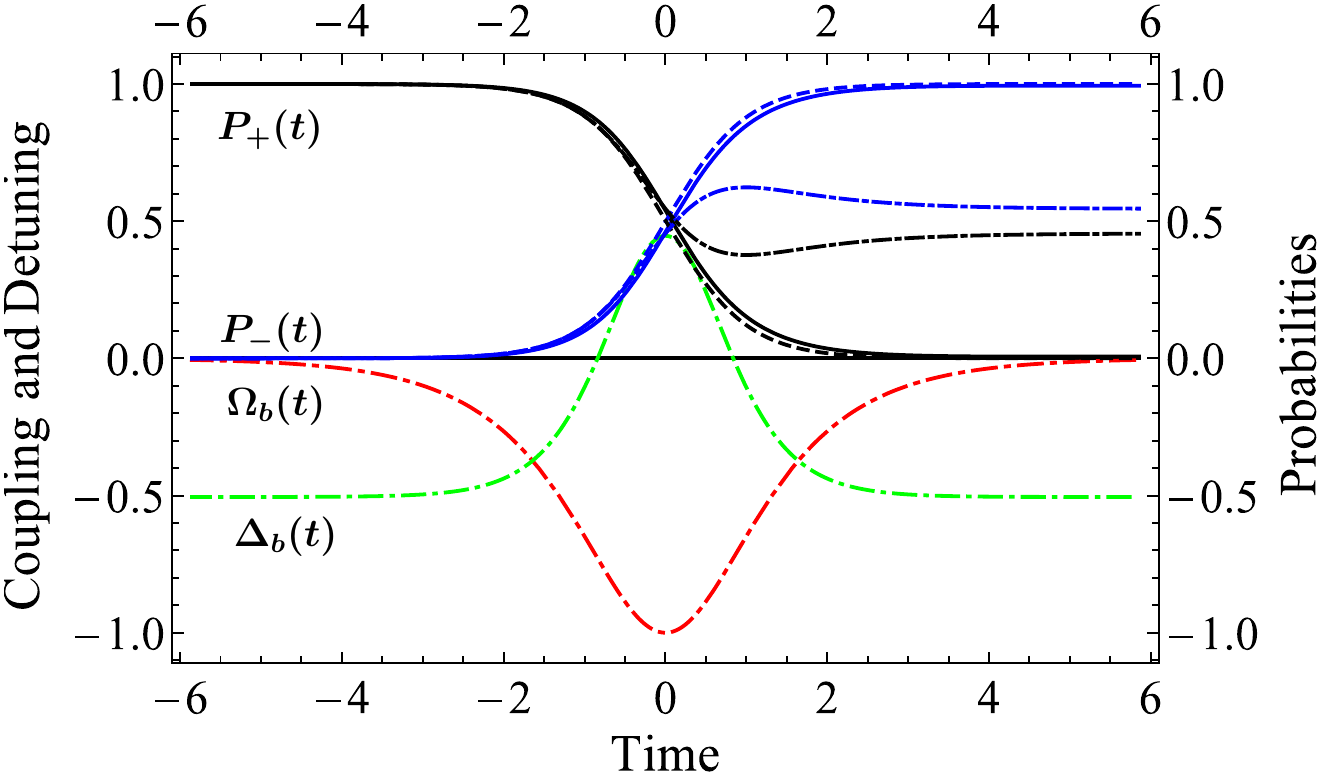}
\caption{
Effective resonant qubit switching dynamics. \textit{Upper panel:} Probabilities for the
qubit states, numerically obtained by solving
(\ref{Schrodinger_s}), are plotted -- $P_+(t)$
in black and $P_-(t)$ in blue. The solid and dashed
curves depict the exact and approximated via $\left[\sqrt{1-\varphi(t)^2} \approx 1 -
\frac12\varphi(t)^2\right]$ probabilities, with correction determined
by $d_{z1}$ (\ref{coupling_d_z1}). The dash-dotted curves correspond
to the
noncorrected probabilities at $d_{z1}\equiv 0$.
\textit{Lower panel:}
The coupling
$\Omega_b(t)$, $(\Omega_{b0}T=-1)$, and the corrected detuning
$\Delta_b(t)$ are shown with dash-dotted red and dash-dotted green,
respectively. The parameters used for computations are
$S=10$, $L=10\Delta x$, $k=\frac{\pi}{30}$, $\mu=\nu$, $JT=4.783$,
$AT=47.834$, $d_{xy}T=2.243$, $d_{z1}T=0.051$ with
$\eta=1.372$. The dimensionless time $t/T$ is plotted on the
abscissa.
}
\label{Fig:SpinSwitching}
\end{figure}

The dynamics of the probabilities $P_{\pm}(t) = \vert
C_{\pm}[s(t)]\vert^2$ for the case of qubit switching at some
specific values of the soliton parameters is shown in Fig.
\ref{Fig:SpinSwitching}.
The necessity of the correction parameter $d_{z1}$ expressed in \eqref{coupling_d_z1}
is evident from the comparison of the probabilities for the
effective on-resonance model with $d_{z1}\neq0$ and without correction
($d_{z1}=0$),
and the exact probabilities numerically obtained
from {\eqref{Schrodinger_s}}. It is clearly seen that the correction
term successfully cures the deficiency introduced by solely using the static term
$d_{z0}$ in Eq. \eqref{coupling_d_z0}.

It is worth noting that the relative phase between the amplitudes
$C_{\pm}(+\infty)$ in the effective resonance regime is altered and
oscillates with time due to the correction term. In the
non-corrected probabilities the phase of the solution is
time-independent.
The relative phase in the superpositions of qubit states
contains an additional phase of $2\omega_b t$ as a consequence of the phase transformation
(\ref{phase_transformation_gamma}).

\subsection{Dark--soliton drive}
The dark soliton solution of NLSE \eqref{NLSE} that exists for $b_kg_k >0$ ($A>0$ and
$\frac{\pi}2<k\leq\pi$) with the amplitude
taking the same value at both ends of the chain, i.e. $x\rightarrow\pm \infty$ reads
\begin{subequations}
\begin{equation}\label{darky}
\varphi_d(x,t) =  \varphi_{d0} \, {\rm tanh} \left(\frac{x-vt}{L}\right) 
\end{equation}
with amplitude
\begin{equation}
\varphi_{d0} = \frac{1}{L}\sqrt{-
\frac{2J\cos k}{J(1-\cos k)+A}}\,
\end{equation}
and frequency
\begin{equation}
\omega_d = 
\omega_0+ 2SJ\frac{\cos k}{L^2},
\end{equation}
\end{subequations}
where the second term describes the correction to the frequency of the
linear spin wave.
Notice that at $k=\pi$, thus $v=0$, we have a static dark soliton.
It is worth mentioning that spin-wave dark solitons have been
theoretically predicted and experimentally realized
\cite{slavin_bright_1994,slavin_generation_1999,bischof_generation_2005}.

\begin{figure}[h!]
\centering
\includegraphics[width=\columnwidth]{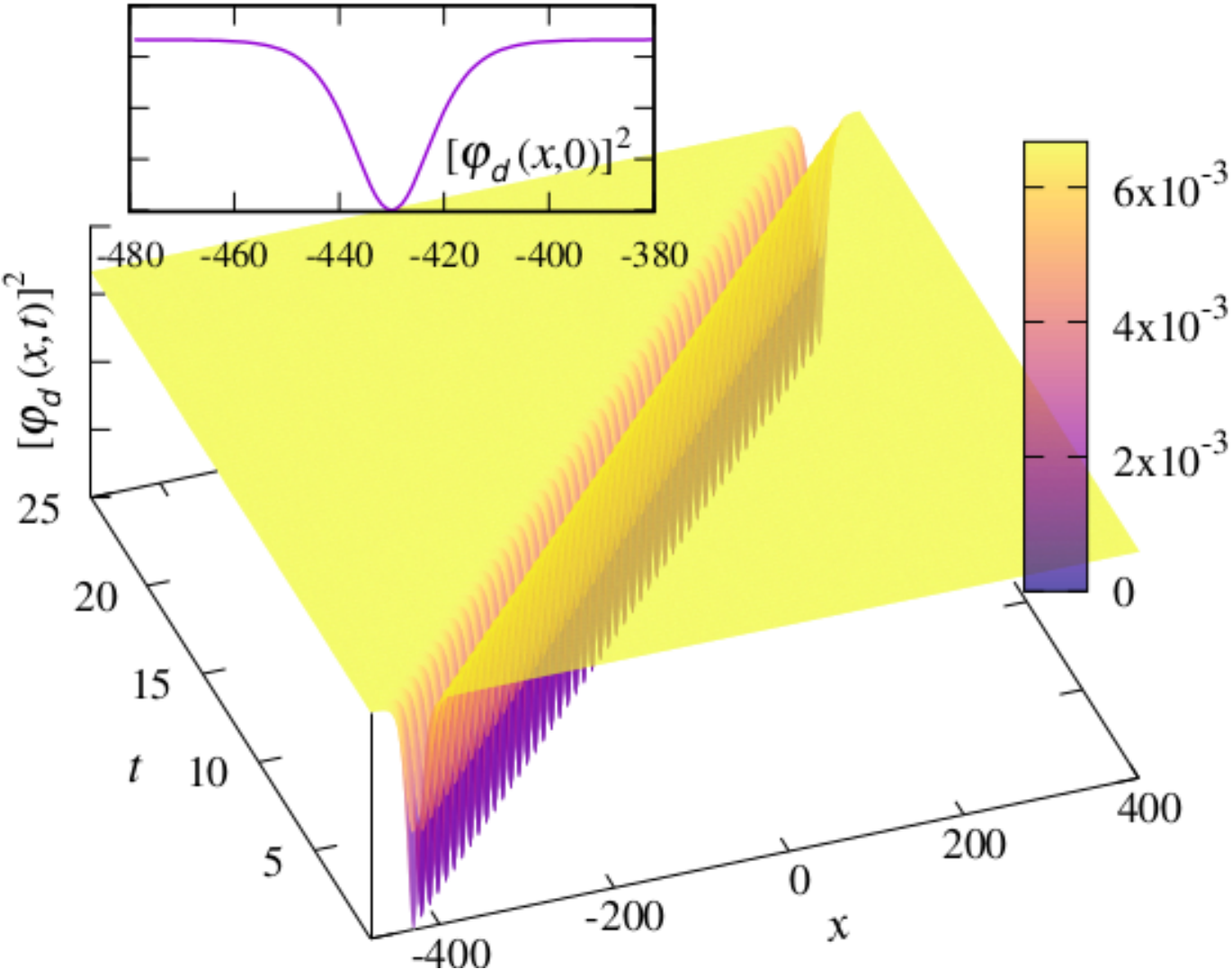}
\caption{Time evolution of the square of the dark soliton solution
$[\varphi_d({x,t})]^2$ in
\eqref{darky} computed with the set of parameters $k = \pi - 0.015708$ and
$L=10$ for $S=J=A=1$. It is clearly seen that the soliton
preserves its shape while propagating along the ferromagnetic Heisenberg chain. The
inset shows the profile of the solitary wave as a function of the
position on the spin chain at the initial time $t=0$.}
\label{darksol}
\end{figure}

We have investigated the evolution of a propagating dark soliton
numerically based on the discrete equations of motion \eqref{EOMa}, and as in the case of bright
soliton, it is generated at time $t=0$ away from the qubit position.
Our numerical simulation have been performed on a chain with 1000
spin sites with $J=S=A=1$ under periodic boundary conditions.
Remark that the dark solitary wave does
not alter during the propagation along the chain (see Fig.
\ref{darksol}).

When a dark soliton propagates in the chain and the qubit is
coupled locally to a spin on the Heisenberg chain, the
variables \eqref{st} and \eqref{thetas}, respectively, read
\begin{subequations}
\begin{equation}
s_d(t) =  {\rm sgn}(t)\,\frac{\Omega_{d0}T}{2} \, \ln\cosh\left(\frac{t}{T}\right)  , \label{s_of_t_DSLI}
\end{equation}
and
\begin{equation}
	\Theta_d(s) =  -  \frac{\Delta_{d0}}{d_{xy} S} \,\,
	\tilde{\varphi}_d(s)^{-1}
- \frac{d_z }{2d_{xy}} \,\, \tilde{\varphi}_d(s) ,  
\label{Theta_DSSSQ}
\end{equation}
\end{subequations}
where the amplitude of the coupling \eqref{Omega_def} now reads
$$
\Omega_{d0} = d_{xy}S\varphi_{d0},
$$
while the time-independent term of the detuning \eqref{delta_def} is given
by
$$
\Delta_{d0} = 
(\mu-\nu)H_0-d_zS+\omega_d
$$
and
$$
\tilde{\varphi}_d(s)
= - \varphi_{d0} \left[1 - \frac{e^{-2f(s)}}{1 + {\rm sgn}(s) \sqrt{2e^{-f(s)}\sinh{f(s)}}}\right],
$$
with
$$
f(s) = \frac{2}{\Omega_{d0}T}\ {\rm sgn}(s) \ s.
$$

In order to obtain an invertible transformation between time $t$ and
the variable $s$ \eqref{st} on the entire real axis, we used the
sign function in (\ref{s_of_t_DSLI}). Moreover, it can be easily seen
that this transformation remains smooth.

Similar to the case of bright soliton, the qubit is set at the
position $x=0$, and the square root in the detuning \eqref{delta_def} 
is expanded in
Taylor series with respect to $\tilde{\varphi}_d(s)^2$. The second term
in (\ref{Theta_DSSSQ}) is of order $O(\varphi^2_{d0})$ compared to
the first one, and can be neglected given that
the counterpart of (\ref{delta_large_BS-SSD}) for the dark soliton is
fulfilled. Then, the problem reduces to the study of a model with hyperbolic tangent coupling
\begin{subequations}
\begin{equation}
	\Omega_d(t)=\Omega_{d0}\tanh\frac tT,
\end{equation}
and constant detuning
\begin{equation}
\Delta_d(t)=\mathrm{const.}
\end{equation}
\end{subequations}
This is the so called ``$\tanh$'' model, that is exactly solvable
and its physical properties are well known \cite{Simeonov2014}.
In the following, we will take advantage of the analysis of Ref.
\cite{Simeonov2014} to gain insights into the effect of the dark
soliton on the behavior of the qubit. Let us point out that the model
considered in Ref. \cite{Simeonov2014} is restricted to positive
times, while here we extend the study over the whole real time axis.

The probability $P_-(t)$ of the qubit flip may take any value in the interval $[0,1]$
for suitably chosen control parameters. For
instance,
it is possible to achieve a return of the
qubit to its initial state. Furthermore at effective resonance,
$\Delta_d=0$, both
a return to the initial state and a qubit switch are possible.
We will focus on the limiting cases of fast $\left(\frac tT\gg1\right)$ and
slow $\left(\frac tT\ll1\right)$ solitons reflected by the behavior
of the coupling's tanh function and resonance.
These are
of particular interest since they provide analytical closed-form solutions
in terms of some elementary mathematical functions.
For further consideration, we assume that the
interaction time of the soliton with the qubit is symmetric with respect
to the reference time $t=0$, except for the resonance case.

The fast soliton is characterized
by a ${\rm sgn}(t)$-dependent constant coupling
\begin{equation}
\label{T_small_DSSS}
\Omega_d(t)\sim\tanh\left(\frac{t}{T}\right) \approx {\rm sgn}(t), \quad t\gg T. 
\end{equation} 
The large--$S$ assumption ensures the
fulfillment of the above condition that holds for a
large set of combinations of the soliton's parameters.
After some lengthy, but straightforward computations,
under the assumptions $|\Delta_b T| \ll 1$, $|\Omega_{b0} T| \ll 1$,
$\Omega_{d0}
\sim \Delta_{d}$ and \textit{for some specific values of $t$}, the leading behavior of the transition probability takes the form
\begin{equation}
P_-(t) \sim P^0_- \left[ \sin^2 \, \frac{\chi(t)}{2} + \frac{1}{4} \sin^2 \chi(t) \right], 
\label{P_-_DSSS_P_ti_eq_t_ad_small}
\end{equation}
where the amplitude and the phase read
$$
P^0_- = \left(2\frac{\Omega_{d0}\Delta_d}{\Omega_{d0}^2+\Delta_d^2} \right)^2 \sim 1,
$$
and
$$
\chi(t) = - \Omega_{d0} t,
\label{DSSS_P^0_-_adsmall} 
$$
respectively.

A peculiar feature of the transition probability $P_-(t)$ is that 
it consists of the superposition of two oscillating time-dependent
components possessing different frequencies and
amplitudes $\chi(t)$ and $\frac12\chi(t)$. Eq.
(\ref{P_-_DSSS_P_ti_eq_t_ad_small}) implies that this
behavior provides the opportunity to tune $P_-(t)$ to a specific
preselected value, even close to unity.
In this approximation, the transition probability $P_-(t)$ tends to the Rabi
model probability with a jump through change of the sign of the
coupling.

In the regime where $|\Delta_d T|$ is finite and $|\Omega_{d0} T| \ll 1$, we have $P_-(t)
\propto \left(\frac12\Omega_{d0} T\right)^2$, and hence such regime is not of interest, when one
needs to achieve an appreciable qubit flip probability.

Let's point out, that the case $|\Delta_d T| \ll 1$ and
finite $|\Omega_{d0} T|$
related to the zeroth order approximation of $P_-(t)$ is contained in
the general resonance solution (\ref{Resonance_DSCS}), which vanishes
in a symmetric time interval.

The limiting case $T \gg t$ is typical for a slow
soliton, and hence, by a linear time-dependent coupling:
\begin{equation}
\Omega_d \sim \tanh\left(\frac{t}{T}\right) \approx \frac{t}{T}, \quad t\ll T.
\label{DSSS_T_Large}
\end{equation}
This behavior shows up as $k \rightarrow \pi$. Then, the tanh model
may be mapped onto the Landau-Zener model rotated by $\frac\pi4$ that
has been investigated in details in Ref. 
\cite{Torosov2008}.

We consider the regime of sufficiently
large coupling amplitude, i.e.
$\Omega_{d0}\ t^2 \gg T $
and the two limiting cases: $|\Delta_d T| \ll 2\Omega_{d0}t$
corresponding to a small detuning regime
and $|\Delta_d T| \gg 2\sqrt{\Omega_{d0}T}$ associted to the large detuning
regime. Here, the transition probabilities $P_-(t)$ are oscillatory and
can be controlled to some extent by the soliton's tunable parameters.
Let's point out that the case of large detuning case corresponds to the adiabatic solution.
On the other hand, the case of weak
coupling regime $\Omega_{d0}\ t^2 \ll T$ is not of interest, due to the negligibly small
qubit flip probability.

Finally we will turn our attention to the
on--resonance ($\Delta_d = 0$) regime, when the variable
(\ref{Theta_DSSSQ}) reduces to
the second term only. Then, the detuning, $\Delta_d(t) \propto
\vert\varphi_d(t)\vert^2\ll1$, may be neglected, since it is almost zero in the
region where the tanh-shaped coupling changes its sign, and it is
vanishingly small elsewhere. In this effective on-resonance
regime the transition probability reads
\begin{equation}
	P_-(t) \approx \sin^2 \left[\frac12\Omega_{d0}T \, \ln\,
	\frac{\cosh\left(t/T\right)}{\cosh\left(t_{\rm i}/T\right)}
	\right].
\end{equation}
It can be well approximated by the resonance Rabi model solution 
\begin{equation}
	P_-(t) \approx \sin^2 \left(\frac{\Omega_{d0} \tau}{2} \right)  , \,\,
t, t_i \gtrsim T, \,\, \tau = t - t_{\rm i}. 
\label{Resonance_DSCS}
\end{equation}
Indicating that, the evolution of the qubit undergoes slightly chirped
oscillations with almost unit amplitude. Therefore, a qubit switch
($\Omega_{d0}\tau = (2p+1)\pi, \,p\in\mathbb{Z}$) and a return
($\Omega_{d0}\tau = 2p\pi, \,p\in\mathbb{Z}$) of the qubit to its
initial state, and
superpositions of qubit states are possible to achieve for
some specifically chosen controlable parameters. In particular, the choice
of a symmetric time interval $t_i = t$ would always lead to a return of
the qubit, because in resonance an anti-symmetric coupling
function in symmetric interval produces $\Omega_{d0}\tau=0$.

\section{Conclusions}\label{Sec:Discussion_and_Conclusions}
Manipulating the state of a qubit is crucial to quantum
computing and quantum information processing. This may
be achieved with the aid of an external stimulus. Here we
consider the action of an anisotropic Heisenberg spin chain on a qubit placed a
distance apart to get rid of the effect of decoherence (information
loss) due to the backward
effect of the qubit on the spin chain.
To achieve our goal, we work in the
large-spin approximation that allows us to map the original chain --
qubit Hamiltonian onto a two--level problem of a qubit under the
action of a propagating through the chain magnetic solitary wave.
Thus, we may take control over the qubit state by tuning the soliton
parameters. Here, we demonstrate the possibility to control the qubit
state through a bright soliton or its dark counterpart. It is shown,
among other, that the qubit can be flipped and/or returned in its
initial state, and an equal superposition of qubit `up' and
`down' states can be generated.

In the case of a local interaction of the qubit with its closest spin on the
chain, the off-resonance and effective resonance regimes are studied.
In all cases the considered problem is
mapped onto some exactly solvable models. To achieve an effective
resonance regime for a bright soliton a fine tuning of the
$z$-component of the qubit coupling is required, while in the case of dark soliton
this is not needed due to the smallness of the coupling
around the time origin.

Finally, we believe that such a scheme for control of qubit by
solitons may find application in systems, such as magnetic chains coupled to an
artificially designed effective half-spin or a coherent atomic spin
chain coupled to an artificially designed effective qubit.

\acknowledgments

The authors would like to thank Prof. T. Mishonov for helpful
discussions.
This work was supported by the Bulgarian National Science Fund under
grant No K$\Pi$-06-H38/6 and the National Program ``Young scientists and
postdoctoral researchers'' approved by  PMC 577 / 17.08.2018.


%
\end{document}